
\nopagenumbers
\magnification=\magstep1
\hsize=5.75truein
\vsize=8.75truein
\centerline{PARTICLE ACCELERATION IN (BY) ACCRETION DISCS}
\bigskip
\centerline{J. I. Katz}
\medskip
\centerline{Department of Physics and McDonnell Center for the Space
Sciences}
\centerline{Washington University, St. Louis, Mo. 63130}
\bigskip
\centerline{ABSTRACT}
\bigskip
I present a model for acceleration of protons by the second-order Fermi
process acting on randomly scrambled magnetic flux arches above an accretion
disc.  The accelerated protons collide with thermal protons in the disc,
producing degraded energetic protons, charged and neutral pions, and
neutrons.  The pions produce gamma-rays by spontaneous decay of $\pi^0$ and
by bremsstrahlung and Compton processes following the decay of $\pi^\pm$ to
$e^\pm$.
\bigskip
\centerline{INTRODUCTION}
\bigskip
The most remarkable property of AGNs is the appearance, in many cases, of much
of their luminosity as the acceleration of nonthermal particles.  Evidence
for this consists of the polarized optical continuum with a power law
spectrum found in some AGN, the great power required to supply radiating
electrons to radio galaxies, and the gamma-ray luminosity of $10^{48}$
erg/sec of 3C279 recently discovered$^1$ by GRO.

This paper is a preliminary account of the calculation of a model$^2$ of
particle acceleration in low density astronomical shear flows.  Particles
are accelerated by a second-order Fermi mechanism.  They are assumed to be
trapped in magnetic mirrors consisting of magnetic flux arches whose feet
are pinned to the surface of a quasi-Keplerian accretion disc.  The
differential motion of points on the disc surface at differing radii
accelerates the particles.  The resulting forces on the disc (acting through
the magnetic field) are described by a viscosity, and may be its chief
dissipative process.  This mechanism directly converts the gravitational
power of black hole accretion to particle acceleration.  Although this
process is second order in the Keplerian velocity, in rapidly rotating inner
discs it may be rapid.
\bigskip
\centerline{CALCULATION}
\bigskip
The greatest uncertainty of this model is the magnetic field configuration,
particularly how flux tubes connect disc points at different radii.  This
uncertainly is not presently resolvable empirically or theoretically.  In
addition, it is not known how well the magnetic arches act as mirrors, nor
is the density distribution of thermal gas within and above the disc known.
The accelerated protons collide with the gas
(the interaction of GeV protons with radiation is negligible, but Compton
scattering prevents the acceleration of electrons).  These uncertainties
affect both the acceleration rate and the loss rate, and may be combined
into a single parameter describing the ratio between these two rates.

The evolution of the isotropic volume-averaged proton distribution function
in momentum space $n(p)$ is given by
$${\partial n(p) \over \partial t} = {\vec \nabla}_p \cdot [D(p) {\vec
\nabla}_p n(p)] - \rho \sigma(p) v(p) n(p) + \rho \int v(p^\prime)
\sigma(p^\prime,p) n(p^\prime)\, dp^\prime; \eqno(1)$$
on the right hand side the first term represents the momentum space
diffusion, the second represents collisional losses in thermal matter of
density $\rho$, and the third represents the contribution of collision
products to the proton distribution.  The velocity $v(p) = p/\sqrt{m_p^2 +
p^2/c^2}$ and $\sigma(p)$ and $\sigma(p^\prime,p)$ are total and
differential proton-proton cross-sections.  If the momentum-space diffusion
results from scattering by magnetic mirrors with uncorrelated speed $u$
and has a mean scattering length $\ell$, then
$$D(p) = {4 \over 3}{u^2 \over \ell c^2} pE, \eqno(2)$$
where the coefficient has an order-of-unity uncertainty resulting from the
unspecified correlation between the directions of the incident and scattered
particles.  The parameter $\rho \ell$ describes the comparative importance
of acceleration and collisional losses.

The scattering processes are
$$\eqalignno{p + p &\to p + p &(3a)\cr
p + p &\to p + p + \pi^0 &(3b)\cr
p + p &\to p + n + \pi^+ &(3c)\cr
p + p &\to {\rm others}. &(3d)\cr}$$
A characteristic energy scale is set kinematically by $m_p c^2 = 938$ MeV,
and also by the increase and saturation of the cross-sections for the
inelastic processes (3b) and (3c) in the range 400--700 MeV.  At laboratory
energies $> 1$ GeV (3d) begins to replace (3b) and (3c), but was not
included in these preliminary calculations (although the correct total
inelastic cross-section was used).

The calculations reported here consisted of the evolution of equation (1),
using (2), and the calculation of the spectrum of $\pi^0$ produced in (3b)
and of the subsequent decay gamma-rays.  The spectrum of $\pi^+$ was not
calculated (although the cross-section of [3c] was included in equation [1]);
this important process (its cross-section is about five times that of [3b]),
as well as the subsequent decay $e^+$, and the effects of (3d) are presently
being added to the code.  The differential cross-sections used were scaled
from the measurements of Bugg, {\it et al.}$^3$ at 970 MeV; this is one of
the few experimental papers which give the laboratory distribution of the
energies of the scattering products (most of the literature gives
center-of-momentum energies, which are insufficient unless the scattering
angle is also known).
\bigskip
\centerline{RESULTS}
\bigskip
The effect of (3a) is to multiply the number of energetic protons, while
conserving their total kinetic energy.  Processes (3b)--(3d) have a similar
effect, although some kinetic energy is lost.  The fate of the neutron
produced in (3c) depends on the dimensions of the acceleration region; in
regions larger than $10^{13}$ cm (appropriate to a massive black hole in an
AGN) it decays and may be regarded as equivalent to a proton, while
neutrons are lost from smaller regions (galactic X-ray sources).

Momentum space diffusion multiplies the energetic proton energy, while
reducing their number (some diffuse to a Coulomb drag sink at zero
momentum).  The combination of diffusive acceleration and collisional loss
will, for suitable values of $\rho \ell$, lead to an exponential (in time)
runaway in proton number and energy density, with a stationary normalized
spectrum.  In reality such a runaway would saturate because the growing
particle energy density would disrupt the confining magnetic field, or
because the growing viscosity would deplete the accretion flow.

The following figures show the results of such a calculation.  Particles up
to 6 GeV were included, but cross-sections above about 1 GeV were inaccurate
because of the exclusion of (3d).  The calculated maximum in the gamma-ray
spectrum at $m_{\pi^0}c^2/2$ was not observed in the data$^1$.  This may
perhaps be explained by the contribution of gamma-rays from $e^\pm$
bremsstrahlung, as is the case for the Galactic gamma-ray spectrum.
These processes will be included in the future.
\bigskip
\centerline{REFERENCES}
\bigskip
\noindent
1. R. C. Hartman, {\it et al.}, {\it Ap. J. (Lett.)} {\bf 385}, L1 (1992).
\par
\noindent
2. J. I. Katz, {\it Ap. J.} {\bf 367}, 407 (1991).
\par
\noindent
3. D. V. Bugg, {\it et al.}, {\it Phys. Rev.} {\bf 133}, B1017 (1964).
\par
\vfil
\eject
\pageinsert
\vfil
\vskip6.3truein
\vfil
\centerline{Figure 1: Proton spectrum.}
\endinsert
\pageinsert
\vfil
\vskip6.3truein
\vfil
\centerline{Figure 2: Neutral pion spectrum.}
\endinsert
\pageinsert
\vfil
\vskip6.3truein
\vfil
\centerline{Figure 3: Gamma-ray spectrum from neutral pion decay.}
\endinsert
\bye
\end